# The INFN Experience in Supporting and Improving HEP Outreach

## Giorgio Chiarelli[*]

*a Istituto Nazionale di Fisica Nucleare, Sezione di Pisa,*
*Largo Bruno Pontecorvo, 3, Pisa, Italy*

*E-mail:* giorgio.chiarelli@pi.infn.it

INFN is recognized as an Italian excellence in science. Born in 1951, over time it created a world-wide network of activities spanning from high-energy physics at the most powerful accelerators, to the search for Dark Matter and rare events in deep underground laboratories, flanking the operations of four national laboratories in Italy. However, until a couple of decades ago, its role was not adequately appreciated by the Italian public at large. Since the beginning of the millennium INFN unfurled a strategy aimed not only to promote its image, but also to improve the transfer of knowledge acquired in its operation to different actors. In this paper we will deal with the improvement of outreach, presenting the strategies pursed, and some of the paths followed to this aim. Due to space limitations only a schematic view of a twenty-something years of work will be presented. In the conclusion we will report the results of an external evaluation of our efforts. In this paper we will not discuss the program of refresher courses for teachers, despite their relevance in our strategy to improve INFN participation in lifelong learning activities for the Italian society. Likewise, despite its centrality to improve INFN capability in the realm of outreach, we will not present the training program aimed to our personnel involved in science communication, and we will not touch the complex activity recently started to assess the impact of INFN communication efforts. Finally, we do not discuss the initiatives and the strategies pursued in 2020-2022 due to the COVID19 pandemia, despite the importance of outreach actions taken by the Institute.

---

[*]Speaker

## 1. Introduction

INFN was born in 1951, as a research institution with a mission: "perform research in nuclear and sub-nuclear physics and development of the related technologies". It must also "promote and provide scientific education and engage in diffusion of scientific culture." To this aim we operate four national laboratories, twenty divisions working in close collaboration with Physics Departments, and two national centers. INFN personnel is composed by its own staff (researchers, technologists, technicians, and administrative support), as well as by associated researchers (university staff, students, post-docs etc). Overall, it is a community of about 4,000 people working together to perform the best research in Italy and abroad.

Until the new millennium INFN was relatively unknown to the large Italian public despite the broad spectrum of its research, the national laboratories on Italian soil, a strong participation in CERN research program, and international collaborations with institutions from all over Europe as well as from United States, China, Japan, Russia etc. The first step was the creation in early 2000 of a Communication Office (CO) that, since the beginning, pursued both the institutional aspects of communication and the outreach of INFN research activities. World year of Physic in 2005 was the first time in which both the Communication Office and researchers through Italy developed organized actions to popularize physics in a structured way. As result of this long-term commitment, the situation is now changed[1].

## 2. A New Approach

Since late nineties of the previous century, growing criticisms toward a top-down approach to scientific dissemination to fill a gap (a.k.a. the deficit model) showed that direct engagement of the public with scientists was key to transfer knowledge. At the same time the evolution of the internet changed the paradigm of interaction between producers and customers. The "direct-to-consumer" retail model can be reversed: as the customer needs information, this can be obtained directly from producer, skipping the role of mediators. In a similar fashion, public enjoys and appreciates more the role of scientists, not only as producer of new knowledge, but also as disseminators. Nothing new, but the availability of new media enhanced this role. As an example, a survey of the Italian population showed that the credibility of websites of research institute grew from 47.9% in 2008 to 66.4% in 2012. Over the same period of time blogs of researchers went from 44.9% to 63.1%, public conferences of scientists from 65.4% to 72.4%, while scientific inserts of daily newspapers (as case in point of mediated information) stayed flat at 55% [1].

Since the beginning the CO dealt with the changing paradigm of science communication using a multi-faceted approach, at the same time its contribution to outreach has been key to improve INFN capability to illustrate its research, experiencing many (sometimes unusual) ways to describe our topics to the various publics. As an example, INFN magazine became a powerful tool targeting mainly High School pupils and teachers. With its monothematic issues, an engaging graphics, and a board composed of researchers, *Asimmetrie* is a unique editorial product. Contributions are written by INFN personnel and edited with the support of CO for clarity. From the point of view of combining the institutional communication with dissemination of research, the Higgs discovery represents a model and functioned as a catalyst, at a time when a new awareness of the importance of transferring the knowledge gained in our activities to society was growing within the Institute, as well as the importance to directly involve our researchers. In this circumstance our CO ideated and organized many public events aimed to explain the discovery and its relevance for the layperson. Direct participation of top INFN management as well as of researchers who played major roles in the experiments, was key to success becoming a model[2]. In a similar fashion the success of less academic formats like performances with scientists' interviews interplayed by readings and music, showed to our personnel that time was ripe for a change in approach to popularization.

---

[1] Thanks to CO, the number of citations in press and media doubled in 2015-2019 reaching 9,000 before COVID.

Our management took a second crucial step with the creation (2012) of a National Committee for Technology Transfer (CNTT). Several targets were pursued: provide internal support on exploitation of Intellectual Property (IP), legal and technical support in dealing with external firms, raise awareness of the importance of knowledge transfer within our community, transfer technology to the market. Thanks to the joint efforts of management, CNTT, and a network of local contacts TT became a recognized activity, improving INFN visibility.

The next step to be taken was to create a similar structure to improve, support, and organize outreach activities put forward by our personnel. The last push to this aim was external. The Italian National Agency for Evaluation of the University and Research System (ANVUR) periodically implements an exercise (VQR) to evaluate research. Evaluation of third mission activities evolved: exercise 2004-2010 was a first attempt to list them, the second (2011-2014) performed a test evaluation. Activities were separated in two broad areas: the ones with and the ones without fund transfer towards research institutions. Within the second type, actions were split in diverse topics, the ones about outreach were evaluated in a fully peer-reviewed format. Each Institution provided case studies both at central and local level. INFN scored very well in the former, while locally organized ones showed large room for improvement (almost a factor two difference). In the area of lifelong learning, VQR found INFN not fully exploiting its capabilities. In the wake of this result INFN created a national commission to "coordinate, organize, supervise, and record" third mission activities. CC3M was born.

**2.1 The National Commission for Third Stream Activities (CC3M)**

Born in late 2016, CC3M has the following membership: three researchers, a member of INFN Executive Board, the head of the outreach activities of the Communication Office, the head of the CNTT, one expert on administrative issues, one expert on external funds. This composition ensures a coordination with the management, optimal synergy with the CO, and is capable to financially support the initiatives with its own budget. The initial phase was devoted to audits of existing initiatives, and selection of the ones to be supported. It was also an opportunity to develop guidelines: we require that INFN be noticeably visible both in terms of scientific content and as actor, and support only initiatives with national impact (or at least involving several structures). Proposals must clearly state aim(s), expected impact, resources needed, milestones, and introduce instruments to perform an ex-post evaluation. Activities are evaluated against strategic goals, such as broadening and diversifying our public, tackle the gender gap in STEM, increasing and strengthening lifelong learning programs.

---

[2] A similar approach was taken for the discovery of gravitational waves. For these two cases see G. Chiarelli, A.Varaschin contribution to EUPRIO 2016 Conference.

In the first phase we also built a network of local contacts, one per structure, with the intent of creating a community with shared goals and terms of reference, as well as to be able to collect suggestions and ideas in a bottom-up fashion. This phase was also marked by several topical workshops to discuss pro and cons of previous experiences.

By 2018 we had three pre-existing activities stabilized (RadioLab, the IPPOG Masterclasses, the website Scienzapertutti), and our portfolio was quickly expanding (see Table I reporting current situation) thanks to an independent budget, allocated following established INFN procedures. In a yearly meeting the whole community of CC3M (that is the Commission itself, and the local contacts) debates the activities. Each one has two internal referees that, prior to the meeting, discuss with proponent requests, goals, milestones, and bring in front of the full assembly their suggestions. In this way we implement a transparent approach that comparatively allocates existing resources. At this stage we assign a fraction of the budget, to be able to timely seize the opportunities that might arise. One example is the INFN participation to the Turin Book Fair (the most important one in Italy that attracts more than 150,000 people in a five-day event). In 2018 organizers invited us, as publisher of *Asimmetrie*, on short notice (less than two weeks). Thanks to the effort of our personnel, and to the possibility to quickly allocate the needed funds, we were able to accept and organize a successful participation. In this way, since then, we are present in this Fair with our stand (several thousand visitors per year) for display of our editorial products and perform collateral activities for the public in synergy with other CC3M activities.

**2.1 Broadening our Public**

In 2017 we estimated that INFN activities towards the public "touch" about 60,000 people in a standard year (that is a year in which we do not organize/participate any national exhibition). This figure represents about 0.1% of the total Italian population but we also realized that we always touch the same segments of public. For example, in initiatives towards high schools most participants come from classics or scientific high schools that, overall, represent about only one third of the student population. An analogous situation holds for the public at large, question is: how can we change?

We start from the consideration that there is no such thing as "a" public: there are different publics, each one with its own interests, languages, and use of media. Also, while INFN activities are inspirational and much broader than usually thought, still our research is narrow compared to the whole of physics (and of science). Therefore, we pursue a dual strategy: create initiatives that talk about our scientific themes in dialogue with other disciplines and create/participate in activities with external partners. Indeed, over time the portfolio of initiatives became broader and diversified (see Table I); it is impossible to go through all of them, in the following we will describe some examples.

Art&Science and ASIMOV Prize aim to attract students without a specific interest in science. In A&S students from all over the country conduct a two-year long path that brings them across science and art through lectures, visits to laboratories and museums. They are required to gather in small groups and develop a project for an artistic artifact linked to a given scientific topic. Exhibitions at regional level display those works, and the best five from each are sent to the national final held in Naples. The finalists are on display at MANN (Museo Archeologico Nazionale di Napoli), one of the most prestigious Italian museums. A jury, in turn, selects the best artifacts and the winners are awarded with a one-week stage at CERN. More than 5000 students, from 213 schools (73% scientific and classics) participated to the third edition (2020-2022), producing 1200 artifacts, on display in 13 exhibitions[3].

The other case is the ASIMOV Prize for scientific popularization books. Every year a national committee selects five books among the ones published over the last 2 years, high school pupils review and score these finalists, and in this way, they choose the final winner of the Prize. Students' reviews are (in turn) judged, the best ones declared winners and presented in regional events. ASIMOV Prize started in 2016 with several hundred reviews from five regions, and now (2022 edition) we collected more than 12,000 writeups from participants distributed over twenty regions. The format attracts the same number of male and female students (the second group provides 65% of the winners) and is appealing also to the students with little interest in science.[4]

Another way to tackle the problem of diversifying our public is to partner with other scientific organizations and/or participate in events organized by others. From this point of view a very important role is played by the Communication Office that organizes high-visibility events (like the ones mentioned about the Higgs discovery or the one held in Bologna during ICHEP 2022) [2] and coordinates participation in science fairs (e.g., Festival della Scienza in Genova or Rome Sciences Festival). Over the years the CO ideated several exhibitions: in almost all cases we collaborated with others to broaden the message and, therefore, the interested public. This was the case for *Gravity-Universe after Einstein*, an exhibition on cosmology and gravitation held in 2018 at MAXXI (the Museum of Contemporary Art in Rome) co-organized with the Italian Space Agency, where science was presented in a dialogue with modern art. With more than 150,000 visitors and several collateral events, *Gravity* represents a successful example of what can be achieved. A similar case is *Tre Stazioni per Arte-Scienza* (*Three Stations for Art-Science*), a three-prong exhibition held in Rome in late 2021, where INFN oversaw the station describing *Uncertainty*, Rome University the one about the rise of science in Rome between Restauration and WWII, and Palazzo Nazionale delle Esposizioni took care of the section where contemporary art -inspired by science- was on show.

Before closing this section, we would like to report the opening of a new line of intervention, INFN_KIDS, targeting (mostly) 8-12 years old, a public difficult to reach and that cannot be addressed with usual means. To this aim we are developing a wide variety of activities, from podcast to comics, from hands-on activities (when they can be performed) to online games (like the Advent Calendar for Christmas 2021), and series[5]. Recently INFN_KIDS took part to *Lucca Comics*, a yearly event attracting about 300,000 people in a four-day specialized fair.

## 3. Conclusion

Over the last twenty years INFN developed a strategy to improve in quantity and quality its transfer of knowledge to society. This effort, supported by the creation of appropriate bodies, has successfully evolved from a top-down approach into a dialogue between our researchers and the public at large. CC3M is the last-born of these bodies, with the aim to improve outreach. In VQR 2015-2019 we presented an even number of CO and CC3M (involving local structures) initiatives. Results [3] show the previous gap in quality between central and local initiatives is now closed, also thanks to the constantly increasing synergy between these two bodies. Likewise, we believe that this improvement is present in the many initiatives carried on by our personnel and not coordinated by the Commission.

---

[3] See: *Incorporating creativity and interdisciplinarity in science teaching: The case of Art & Science across Italy*, F. Simone et al., these Proceedings.

[4] See: *The ASIMOV Prize for scientific publishing*, W.M. Alberico et al., these Proceedings.

[5] See *Telling Kids about Physics in interactive online communication formats* C. CollàRuvolo, these Proceedings.

Table I Activities coordinated by CC3M (2022)

| Activity | Typology/periodicity | Target | Contacts |
|---|---|---|---|
| **SaLTo** | Book Fair-yearly | Generic Public | >5000 visitors |
| **Scienzapertutti-SxT** | website | High Schools (HS) pupils, teachers, generic public | >5000 contacts/day |
| **Pint-of-Science** | Talking science in pubs-yearly | Generic Public | Few thousands |
| **Festival di Genova** | Science Fair, yearly | Generic Public | Thousands |
| **ASIMOV** | Literary reviews/prize-yearly | HS pupils Pupils | >12,000 participants |
| **ART & SCIENCE** | Biennial path | HS pupils Pupils | >5000 |
| **OCRA** | Activity with Cosmic Rays, International Cosmic Day, stages | HS pupils Pupils | >1000 |
| **DARK** | various activities (stages, visits) | HS pupils Pupils | >200 |
| **Masterclass IPPOG** | Researcher for one day -yearly (hosted in our structures) | HS pupils Pupils | >3000 |
| **INSPYRE** | International Stage (Laboratori Nazionali di Frascati) | HS pupils from all over the world | About 100 |
| **Lab2Go** | Recovery of HS didactic labs | HS pupils Pupils | About 500 |
| **RADIOLAB** | Radon measurements-yearly | HS pupils Pupils | 500-600 |
| **INFN_KIDS** | Different activities | 8-12 years old | N/A |
| **AGGIORNAMENTI** | Upgrade course (several structures) | Middle School Teachers | About 100/year |
| **PID: Programma Infn per Docenti** | Upgrade course (one weeklong), hosted at LNL, LNGS, LNS | HS Teachers | About 100/year |
| **Incontri di Fisica-IdF** | Upgrade Course (LNF) | HS Teachers | 200/year |
| **Incontri di Fisica Moderna** | Upgrade Course (LNF) | HS Teachers | 20/year |